\documentclass[mathleft]{an}
\usepackage{graphicx}
\usepackage{times}
\usepackage{float}
\usepackage{amssymb}
\usepackage{natbib}
\bibpunct{(}{)}{;}{a}{}{,}
\newcommand{\Ha}{\ensuremath{\hbox{H}\alpha}}
\def\xmm{{\it XMM-Newton}}
\def\cha{{\it Chandra}}
\def\hst{{\it HST}}

\newcommand{\OVII}{\mbox{O\,{\scriptsize VII}}}
\newcommand{\OVIII}{\mbox{{O\,{\scriptsize VIII}}}}

\newcommand{\FeXVII}{\mbox{{Fe\,{\scriptsize XVII}}}}

\newcommand{\Os}{\mbox{{O$^{5+}$}}}

\newcommand{\Lya}{\ensuremath{\hbox{Ly}\alpha~}}
\newcommand{\A}{\AA~}

\begin{document}


\title{On the soft X-ray emission of M82}

\author{Jiren Liu\inst{1}\fnmsep\thanks{Corresponding author:
  \email{jirenliu@nao.cas.cn}\newline},
	Q. Daniel Wang\inst{2},
\and  Shude Mao\inst{1,3}
}
\titlerunning{On the soft X-ray emission of M82}
\authorrunning{J. Liu et al.}
\institute{
	National Astronomical Observatories, 20A Datun Road, Beijing 100012, China
\and 
Department of Astronomy, University of Massachusetts, Amherst, MA 01002, USA
\and 
Jodrell Bank Centre for Astrophysics, University of Manchester,
		       Manchester, M13 9PL, UK}
\received{..}
\accepted{..}
\publonline{later}

\abstract{We present a spatial analysis of the soft X-ray and 
\Ha\ emissions from the outflow of the starburst galaxy M82. 
We find that the two emissions are tightly correlated on various scales. 
The \OVII\ triplet of M82, as resolved by X-ray grating observations 
of \xmm, is dominated by 
the forbidden line, inconsistent with the thermal prediction.
The \OVII\ triplet also shows some spatial variations. We discuss 
three possible explanations for the observed \OVII\ triplet, including the charge 
exchange at interfaces between the hot outflow and neutral cool gas,   
a collisional non-equilibrium-ionization recombining plasma, and resonance 
scattering.
}


\maketitle
\section{Introduction}

Galactic-scale outflows (superwinds) from active star-form\-ing galaxies 
represent an important feedback process, which regulates the galaxy 
evolution and recycles the metals and energy produced by massive stars 
and supernovae \citep{Vei05}.
Indeed, \Os\ ions have been detected out to 150 kpc around star-forming 
galaxies by the 
Cosmic Origins Spectrograph through their absorption \citep{Tum11}. The
details of the feedback process, however, can only be studied effectively 
through the X-ray emitting outflows of nearby starburst galaxies. 
The prototype starburst galaxy M82 (located at 3.6 Mpc), with a powerful 
superwind detected on scales up to 10 kpc, 
is an ideal target to study the physical process of superwind.

The soft X-ray emission of the outflow of M82 is found to be spatially 
correlated with the \Ha\
emission \citep{Str04}, which indicates that the soft X-ray emission 
is due to the interaction between the hot outflow and 
the entrained disk/halo cool gas. 
While the shock-heated gas is expected to produce thermal soft X-ray emission, 
the \OVII\ triplet, as detected with \xmm, is found to be dominated 
by the forbidden line \citep{Ran08, Liu11}, which is inconsistent with the 
thermal prediction.

The K$\alpha$ triplet of He-like ions is
a powerful diagnostic that can be used to test the origin of the X-ray 
line emission \citep[for a recent review, see][]{Por10}.
The triplet consists of a resonance line, two inter-combination lines, and a 
forbidden
line. For a thermal plasma in ionization equilibrium, the electron collisional
excitation is efficient and favours the resonance line. 
The fact that the forbidden line is stronger than the resonance line 
suggests that an alternative emission mechanism may play a role in generating 
the soft X-ray emission.

One interesting possibility is the charge exchange (also called 
charge transfer), which occurs at the interface 
between the hot plasma and neutral cool gas. It has been shown that
highly ionised ions in the solar wind can readily pick up electrons from
neutral species around a comet. These electrons, captured in excited states 
of the ions, cascade down and lead to X-ray line emission
\citep[e.g.,][]{Lis96, Cra97}.
The electron downward cascading favours the forbidden line \citep[for a recent
review, see][]{Den10}.

Similar interfaces between the hot plasma and cool gas also exists 
in the case of galactic superwinds. Observations have shown that the superwinds
contain cool neutral and warm ionized clouds/fi\-laments, as well as highly 
ionized hot gas.
In the southwest outflow of M82 (see Fig. 1), molecular H$_2$ and
CO are observed to be well correlated with the \Ha\ emission, thus with X-ray 
emission \citep{Vei09,Wal02}. 
The correlation between the molecular gas and the \Ha\ emission is less 
prominent in the northeast outflow. 

\begin{figure*}
\begin{center}
\includegraphics[height=2.8in]{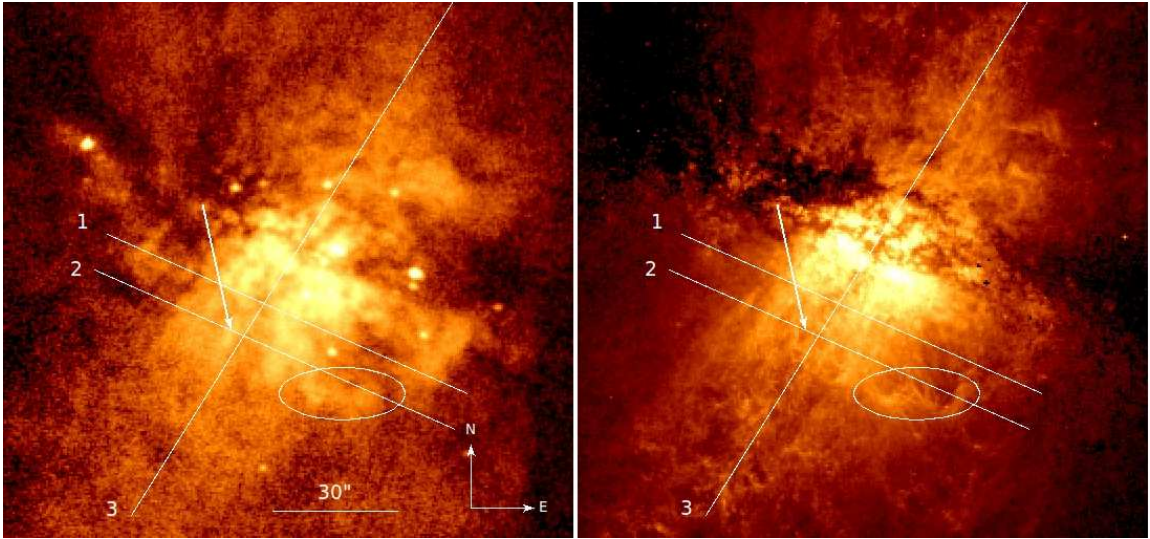}
\end{center}
\caption{Left: \cha\ 0.5-2 keV log-scaled intensity image of M82. Right:
\hst\ continuum subtracted \Ha\ emission of M82. Intensity distributions 
along the three
marked lines are plotted in Fig. 2. The arrow and ellipse mark two regions
that show close spatial connection between X-ray and \Ha\ emissions.
}
\end{figure*}

The relatively strong forbidden line of the \OVII\ triplet of M82 may also 
be produced by the recombination of an overcooled non-equilibrium-ionization 
plasma.
Another process that may have effect on the line ratio is the
scattering of photons of a resonance line, which has a larger optical depth 
due to the large oscillator strength compared to a forbidden line. 

In this paper we study the spatial correlation between the soft X-ray 
and \Ha\ emissions using the high-resolution \cha\ and \hst\ data. We show that 
besides the global correlation between the soft X-ray and \Ha\ emissions, there are 
regions showing that the soft X-ray emission occurs behind the \Ha\ emission 
in the outflowing direction.
We also present a spatially resolved analysis of the \OVII\ triplet of M82.
The three mechanisms mentioned above that can explain a strong forbidden line
are discussed. 

\section{Correlation between the soft X-ray and \Ha\ emissions}

\begin{figure}
\includegraphics[height=4.0in]{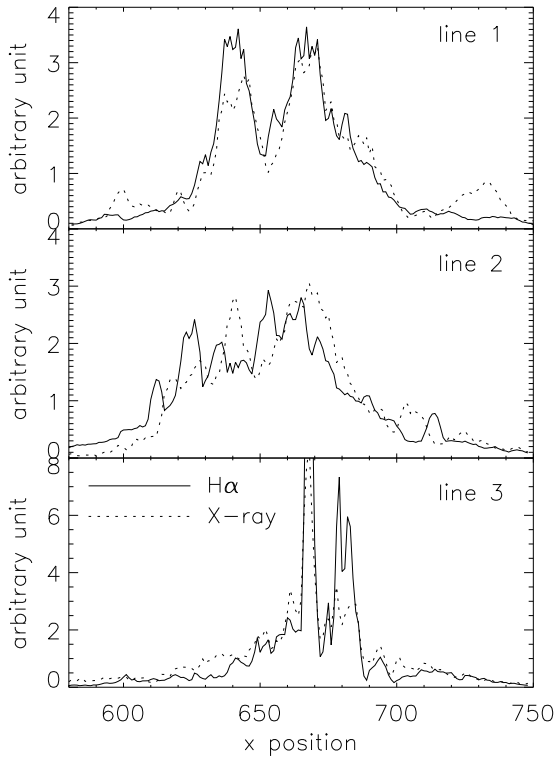}
\caption{The soft X-ray and \Ha\ intensity distributions along three specified lines
marked in Fig. 1. The x-axis is the horizontal pixel index of the image of
Fig. 1 (with 1 pixel corresponding to 0.5$''$).}
\end{figure}

The soft-X-ray emission (0.5-2 keV) of M82 observed by \cha\ is plotted 
in Fig. 1 (left panel) together with the continuum subtracted \hst\ \Ha\ 
image (right panel; Mutchler et al. 2007). The \cha\ ACIS-S3 data are extracted 
from the observations with IDs of 10542, 10543, 10544, 10545, 
10925, 11800 (PI: D. Strickland).
The \hst\ \Ha\ data, binned to 0.5$''$ to match the pixel size of the 
\cha\ image, shows many filamentary structures, as well as loops 
and arcs, while the X-ray image shows limb-brightening features.
We see that in general, the morphology of X-ray emission follows that of \Ha\ 
emission very well, though the \Ha\ image shows much more detailed structures.

To illustrate their relation on finer scales, in Fig. 2, we plot the X-ray 
and \Ha\ intensity distributions along three 
lines marked in Fig. 1. On large scales the profiles of both emissions 
follow with each other, but on small scales, there are regions showing 
distinctive features.
For example, in the region marked by the ellipse in Fig. 1, the X-ray
emission occurs behind the \Ha\ emission in the outflowing direction
(X-ray peak at the position 705 of line 2), and the \Ha\ emission
seems arising from shell-like structures driven by the X-ray emitting
outflow; at another region marked by the arrow, the peak of the X-ray filament
coincides with faint \Ha\ emission (X-ray peak at the position 640 of
line 2).

Because of the projection effect, it is uncertain as to whether the large-scale
similarity of the X-ray and \Ha\ emissions is due to the superposition of 
small-scale adjacent features as shown along line 2 or due to the intrinsic
association of two 
emission components, but the correlation indicates that the soft X-ray emission 
and \Ha\ emission are closely connected.

\section{OVII triplet of M82}

As stated in the introduction, the line ratios of the \OVII\ triplet can be 
used to test the origin of the X-ray emission. To apply the test, the \OVII\ triplet 
should be resolved. With its large dispersion power, \xmm\ Reflection Grating 
Spectrometers (RGSs) \citep{den01} have the unique capability to provide high
resolution spectra for extended sources, such as the superwind of M82. 
We use two exposures with similar observational
configurations (ID 0206080101 and 0560181301) and with a total effective 
exposure of 90 ks. To study the spatial behavior of the triplet, we divide
the cross-dispersion range into three regions: A (-30$''$ -- 30$''$), 
B (-90$''$ -- -30$''$), and C (30$''$ -- 90$''$). Fig. 3 shows the dispersion
direction and extraction regions plotted on the \xmm\ EPIC-pn image of M82.

\begin{figure}
\includegraphics[height=2.0in]{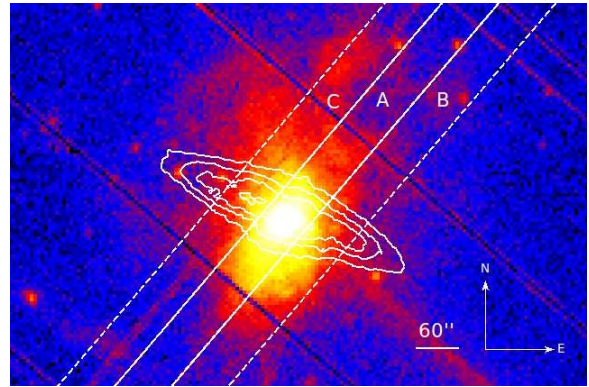}
\caption{Spectral extraction regions over-plotted on the soft X-ray 
(0.5-2 keV) \xmm\ EPIC-pn image of M82. The presented lines outline the RGS 
dispersion direction and the
regions (A, B, and C) from which the RGS spectra are exacted. 
Contours are from the DSS2 red image and indicate the disk of M82. 
}
\end{figure}

The \OVII\ triplets extracted from the three regions are plotted in Fig. 4.
It can be seen clearly that the triplet of the A and B regions is dominated by 
the forbidden line. In contrast, the triplet is dominated 
by the resonance line in the C region.
To study the line ratios of the \OVII\ triplet quantitatively, we fit a
model consisting of three Gaussians and a constant continuum to the data, 
which is expressed as
\begin{equation}
\hspace{-0.5cm}
f=\frac{1}{\sqrt{2\pi}\sigma_{\lambda}}\sum_{j=\rm r,i,f}
f_j\exp\left[-\frac{(\lambda-\lambda_j-\Delta\lambda)^2}
{2\sigma_{\lambda}^2} \right] + f_{\rm con}, 
\end{equation}
where $\lambda_{\rm r,i,f}$ (21.6, 21.8 and 22.1 \AA) are the wavelengths
of the resonance,
inter-combination, and forbidden lines, respectively, $f_{\rm r,i,f}$ 
are the corresponding fluxes, $f_{\rm con}$ is the continuum flux,
$\sigma_{\lambda}$ the dispersion, and  $\Delta\lambda$ the wavelength
shift. The best-fitting results are plotted in Fig. 4 \citep[for details,
please see][]{Liu12}.

\begin{figure}
\includegraphics[height=4.0in]{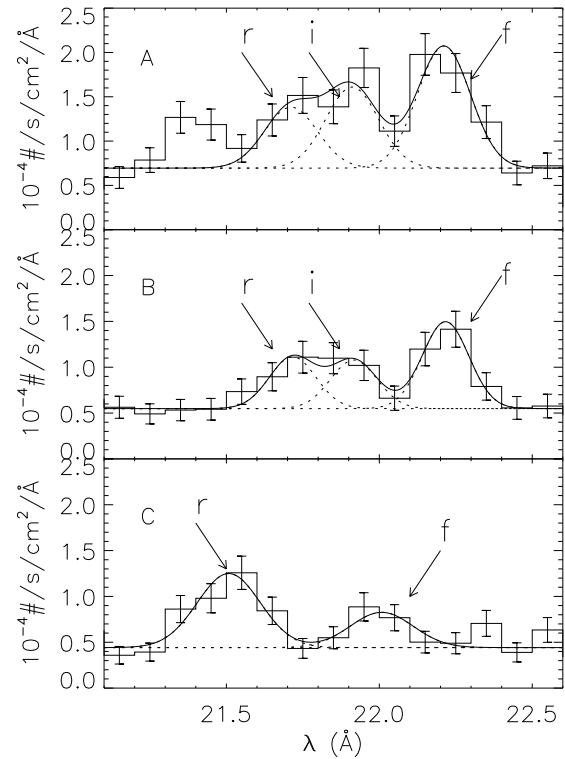}
\caption{RGS spectra of the \OVII\ triplet for the three regions of M82 and 
the best-fit models (solid lines) of three Gaussians (dotted lines) and a constant
continuum. The fitted resonance (r), inter-combination (i), and forbidden (f) lines
are marked with arrows.}
\end{figure}

For the emission from a thermal plasma in ionization equilibrium
and at a temperature greater than 0.1 keV, the ratio 
$G=\frac{f_{\rm f}+f_{\rm i}}{f_{\rm r}}$ is smaller
than 1 \citep[e.g.,][]{Por01}.
The $G$ ratios of A and B regions are around 3, inconsistent
with the thermal prediction.
The $G$ ratio in the C region is consistent with that of a thermal 
plasma around $5\times10^6$ K.

\section{Discussions}

\subsection{Resonance scattering}
\begin{figure}
\includegraphics[height=2.3in]{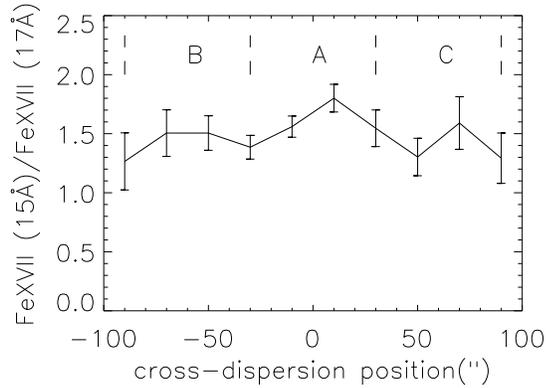}
\caption{The ratio of the \FeXVII\ $2p$--$3d$ (15\AA) to 
\FeXVII\ $2p$--$3s$ (17.1\AA) lines as a function of the cross-dispersion
position. The three extraction regions shown in Fig. 3 are marked. The ratio is 
computed by selecting
counts within a 0.2 \A range centered on each line. As the RGS dispersion 
direction is not perpendicular to the outflow direction (Fig. 3), 
it leads to a wavelength-shift depending on the cross-dispersion position.
When computing the ratio we have corrected this effect. 
}
\end{figure}

One possible explanation of the large G ratio is resonance scattering.
If the optical depth is large enough,
the intensity of the resonance line is re-distributed from the central
optically thick region to the outer optically thin region \citep{Gil87}, 
and the ratio $G$ is increased in the central region and reduced in outer
region. 
For the A and B regions, a diminishing factor of at least 3 is 
needed for the observed resonance line. If the relatively weak resonance line
is indeed due to the resonance scattering, such a diminishment is also expected
for other resonance lines, such as the \FeXVII\ $2p$--$3d$ line at 15\AA. Assuming
the ratio of Fe/O is similar to the solar value and a temperature
around $3\times10^6$ K (estimated from the flux ratio of the \OVIII\ \Lya 
to \OVII\ triplet), the optical depth of the \FeXVII\ 
$2p$--$3d$ line is similar to that of the \OVII\ resonance line. 
Because both the resonance \FeXVII\ $2p$--$3d$ line and the 
optically thin \FeXVII\ $2p$--$3s$ line at 17\AA\ are relatively isolated,
their ratio is a good test of the resonance scattering effect \citep{Xu02}.
In Fig. 5, we plot the cross-dispersion ratio 
of the \FeXVII\ $2p$--$3d$ line to the optically thin \FeXVII\ $2p$--$3s$ line. 
It shows little evidence for the diminishment of the $2p$--$3d$ line 
compared to 
the optically thin $2p$--$3s$ line, which argues against the resonance scattering effect.

\subsection{Non-equilibrium-ionization plasma}

A collisional non-equilibrium-ionization (NEI) recombining plasma
can also produce a relatively strong forbidden line \citep[e.g.,][]{OP04}.
A plasma, if expanding and cooling sufficiently fast, can become overly
ionized. The recombination of the ions and electrons would then dominate the
X-ray line emission. 
However, the X-ray emission of M82 is spatially correlated with the filamentary
H$\alpha$ emission. This fact is consistent with the scenario that the
X-ray emission arises from the interfaces between the hot outflow and the
entrained cool gas, but would be difficult to explain with the NEI model 
in which the emission should come from the bulk of the outflow,
where the low density regions will be the most over-ionized.
Furthermore, the A region shown in Fig. 3 is approximately the star-forming
region, and	the plasma there is hard to cool to low enough temperatures 
that the NEI condition can be achieved.

\subsection{Charge exchange X-ray emission}

The charge exchange X-ray emission is the mechanism that can 
naturally explain the dominance 
of the forbidden line of the \OVII\ triplet. It is supported 
by the tight correlation between the X-ray emission and the cool
gas in the southwest outflow. The thermal-like $G$ ratio of the C region seems 
to arise from the northeast region with faint molecular gas \citep{Vei09}. 
One spectral feature that can be used to further test the charge exchange
X-ray emission is the enhanced 
line flux from the levels ($n=3-6$) of charge-exchange captured electrons,
such as \OVIII\ Ly$\beta$ and Ly$\gamma$ lines \citep{Bei03}.
Another feature to test between the charge exchange and NEI model
is the radiative recombination continuum (RRC), which is expected by the 
NEI model and not by the charge exchange.
Unfortunately, current instruments do not have sufficiently large collecting
area and high resolution to measure such spectral lines. 
With the calorimeter spectrometer of Astro-H planned to launch in 2014, 
one may be able to obtain a spatially resolved map of the \OVII\ triplet, which 
could then be used to constrain the details of the charge exchange process.

\section*{Acknowledgments}
We thank Thierry Montmerle and Rosine Lallement for organizing the CXU workshop 
and for their invitation and hospitality.

\end{document}